# LLM-Mirror: A Generated-Persona Approach for Survey Pre-Testing

**Sunwoong Kim\*, Jongho Jeong\*, Jin Soo Han, Donghyuk Shin**

College of Business, Korea Advanced Institute of Science and Technology (KAIST)

{kingsunwoong, flyeng, jinsoo.han, dhs}@kaist.ac.kr

## Abstract

Surveys are widely used in social sciences to understand human behavior, but their implementation often involves iterative adjustments that demand significant effort and resources. To this end, researchers have increasingly turned to large language models (LLMs) to simulate human behavior. While existing studies have focused on distributional similarities, individual-level comparisons remain underexplored. Building upon prior work, we investigate whether providing LLMs with respondents' prior information can replicate both statistical distributions and individual decision-making patterns using Partial Least Squares Structural Equation Modeling (PLS-SEM), a well-established causal analysis method. We also introduce the concept of the **LLM-Mirror**, user personas generated by supplying respondent-specific information to the LLM. By comparing responses generated by the LLM-Mirror with actual individual survey responses, we assess its effectiveness in replicating individual-level outcomes. Our findings show that: (1) PLS-SEM analysis shows LLM-generated responses align with human responses, (2) LLMs, when provided with respondent-specific information, are capable of reproducing individual human responses, and (3) LLM-Mirror responses closely follow human responses at the individual level. These findings highlight the potential of LLMs as a complementary tool for pre-testing surveys and optimizing research design.

**Code** — https://github.com/anonymous17897234/Mirror
**Datasets** — https://github.com/anonymous17897234/Mirror

## Introduction

Surveys have long been a workhorse methodology for research and decision-making both in academia and industry. They are fundamental tools in various academic fields including but not limited to economics, marketing, political science, and public policy. Hulland, Baumgartner, and Smith (2018) reported that about a third of articles published in prestigious marketing journals employ survey methods.

However, surveys often entail significant time and resource commitments. Even when scholars exert the most meticulous effort in designing questions, surveys are not guaranteed to be error-free. Analyses may yield statistically insignificant outcomes, and non-response rates can lead to selection bias, jeopardizing the validity and generalizability of findings (Groves 2006). While pre-tests or pilot studies can mitigate these issues and enhance the reliability of surveys, they nevertheless require additional resources, further increasing the overall cost of survey implementation.

Given the challenges of conducting human-based surveys, recent advancements in Large Language Models (LLMs) offer researchers a complementary tool for simulating human-like responses. Horton (2023) explored the extent to which LLMs replicate human decision-making in economic scenarios drawn from behavioral economics literature. Argyle et al. (2023) showed that LLMs can predict political outcomes for specific demographic groups. This emerging approach of employing LLMs has promised the potential to mitigate inefficiencies in traditional surveys, such as reducing the need for multiple rounds of pre-testing.

Yet, existing research has primarily focused on whether LLM-generated responses replicate human responses at the *aggregate* level—for example, the *overall* trends in economic decisions in Horton (2023) or the *representative* political preference for certain demographic groups in Argyle et al. (2023).

While these studies highlight LLMs' ability to follow human responses at the aggregate level—evidenced by the alignment between the distributions of LLM-generated and human responses—, they overlook individual-level differences. For example, the average responses of two extreme groups can coincide with the average of a homogenous group, masking the nuanced individual-level variations. The interpretation of the same averages would then differ based on the types of variation that exist in the data. While the average outcome of a homogenous group can be used to infer about the overall population, the same average derived from extreme groups would reflect different channels or mechanisms through which heterogeneous subsets of the population make decisions.

A natural extension of the existing literature is, then, to examine whether LLM-generated responses align with the results of human responses at the individual level. In this paper, we use an LLM to replicate the survey responses reported in Redondo and Aznar (2018) and Damberg, Schwaiger, and Ringle (2022) where the researchers conducted a causal analysis method known as Partial Least Squares-Structural Equation Modeling (PLS-SEM).

---
\* These authors contributed equally.

We first used the LLM to simulate the human responses reported in the existing surveys and then replicated the PLS-SEM analysis using the LLM-generated responses. Specifically, we employed three different approaches based on the information provided to the LLM. The 'Baseline' approach used survey context. The 'Demo prompt' included respondent demographics, and the 'Omni prompt' combined demographics with prior responses that compose the explanatory variable in the PLS-SEM analysis.

Furthermore, we extend our approach by using LLM to generate a 'user persona' based on the respondent's demographics and prior responses. The persona is utilized as the prompt for the LLM to generate responses to the surveys. We defined this process as 'LLM-Mirror', the LLM-driven model of the respondent, designed to closely replicate their attitudes and decision-making processes.

Our work extends prior LLM research by exploring whether LLMs can replicate individual-level responses in surveys. We find that:

- By incorporating survey respondents' demographic information and prior responses related to explanatory variables, LLMs can simulate human responses at the individual level.
- By providing a user persona as a prompt for predicting human responses, the LLM-Mirror can further replicate individual-level human responses.
- Using LLM-generated responses, PLS-SEM analysis produces estimates closely aligned with those using human responses.

Our findings have profound implications for augmenting traditional surveys. When LLMs simulate individual-level human responses with high accuracy, they can be used to test the validity of survey questionnaires and reduce the need for multiple rounds of pre-tests—an essential component in survey research for methodological rigor, albeit the expensive implementation costs.

## Related Work

### LLM as social agent

Recent studies examine how the behavior of LLMs aligns with human behavior (Horton 2023; Tjuatja et al. 2024; Brand, Israeli, and Ngwe 2023; Goli and Singh 2024; Deng et al. 2024; Zhang et al. 2024). Goli and Singh (2024) and Deng et al. (2024) demonstrated that with prompting, LLMs could follow human-like decision-making and rational behavior consistent with economic theory. Tjuatja et al. (2024) demonstrated that LLMs perform well in replicating human opinion distributions as measured by Wasserstein distance. However, they emphasized that this does not mean LLMs can accurately reflect individual behavior. Additionally, Yoo (2024) and Gao et al. (2024) showed that while prompt tuning provides a baseline for following human response distributions, fine-tuning improves LLMs' ability to capture similar outcomes to human behavior. Building on prior studies that examined statistical similarities between the LLM and human decisions, our research uses structural models to explore how information shapes outcomes, uncovering causal relationships and addressing the underlying mechanisms of human behavior, which are essential for analyzing social phenomena.

### PLS-SEM

In social science research, Structural Equation Modeling (SEM) and Partial Least Squares (PLS) are widely employed to examine relationships between latent variables. Chaudhuri and Holbrook (2001) and Bhattacherjee (2001) applied SEM to investigate causal relationships among latent variables, whereas Venkatesh, Thong, and Xu (2012) employed PLS to extend the Technology Acceptance Model. PLS-SEM integrates the strengths of both SEM and PLS, addressing their limitations by enabling the examination of relationships even with non-normal data or small sample sizes. For instance, Choi, Lee, and Yoo (2010) utilized PLS-SEM to analyze the impact of the transactive memory system on knowledge sharing and Hamari et al. (2016) examined the effects of student engagement on learning outcomes using PLS-SEM.

## Methodology

### Experiment Settings

We provided respondents' demographic information along with their answers to prior survey questions, such as those assessing attitudes or preferences. In each study, the survey items were designed to measure specific latent variables. For instance, in Study 1, questions 1 to 4 assessed the "Pleasure induced by online advertising." The latent variables were estimated using the PLS-SEM method based on responses, and the correlations between these variables were analyzed, followed by an examination of the path coefficients.

To assess how the provided information enhances LLMs' replication performance, we tested four types of prompts using the LLM with each differing in the information they provided. The 'Baseline prompting' approach encompassed basic survey contexts, whereas the 'Demo prompting' approach included demographic attributes. The 'Omni-prompting' approach builds upon the Demo prompting by incorporating prior questions and answers.

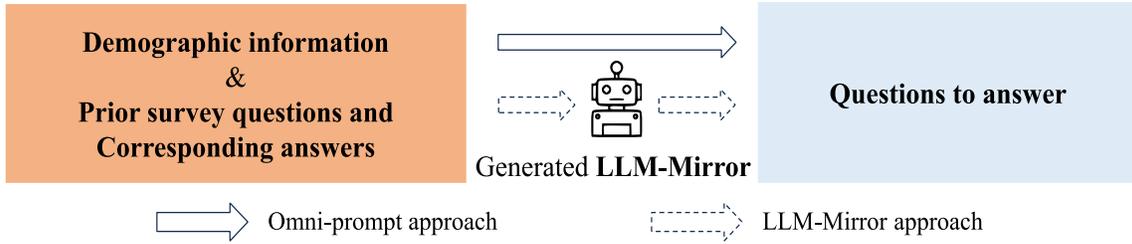

Figure 1: Process of the Omni-prompt approach and the LLM-Mirror approach

Prior questions and answers consist of survey items and responses designed to measure latent factors (underlying variables influencing latent outcomes). Using this information, the LLM evaluates the respondent's attitude, responding to the remaining questions from their perspective. These remaining questions assess the latent outcomes, and in the original study, the correlations between latent factors and latent outcomes were estimated through path coefficients. We compared the path coefficients derived from LLM-generated responses with those based on the original human responses.

The LLM-Mirror approach differs from the Omni prompting approach since it utilizes a user persona instead of detailed prior questions and answers. The LLM-Mirror can reflect an individual's perspective using only a persona, which is a more practical alternative to detailed survey responses in pre-tests. Figure 1 illustrates the process of the Omni prompting and the LLM-Mirror approaches.

All studies were conducted using the GPT-4o LLM model. The details about provided information are summarized in Table 1.

|  | Baseline prompt | Demo prompt | LLM-Mirror | Omni prompt |
|---|---|---|---|---|
| Survey context | O | O | O | O |
| Demographic | X | O | X | O |
| Prior questions and answers | X | X | X | O |
| Generated-Persona | X | X | O | X |
| Questions to answer | O | O | O | O |

Table 1: Details of information provided in each approach

## Evaluation Metrics

We compared the response distributions between the LLM and human responses using Jensen-Shannon divergence (Lin 1991) and Wasserstein distance, based on response frequencies. Additionally, we compared the path coefficients from the PLS-SEM results.

Jensen-Shannon divergence is a symmetric and bounded measure that quantifies the difference between two probability distributions. A value of 0 means the distributions are identical, while a value near 1 indicates they are completely different.

Wasserstein distance is useful for comparing survey responses, those involving continuous or ordered data, as it captures the overall differences in their shape and structure.

In this study, we replicated parts of the structural model used by Redondo and Aznar (2018) and Damberg, Schwaiger, and Ringle (2022) to estimate path coefficients. We employed the *plspm* library in R, using 5,000 bootstrap samples in Study 1 and 10,000 bootstrap samples in Study 2 to match the original study's setup. The original studies utilized *SmartPLS* for PLS-SEM analysis. Given potential differences in estimation methods between *SmartPLS* and *plspm*, we conducted our own analysis for human responses. We then compared the path coefficients derived from human responses with those obtained from the LLM responses.

To further validate the consistency between LLM and human responses at the individual level, we grouped responses into categories of disagreement (1-3), neutrality (4), and agreement (5-7). This analysis evaluated how closely LLM responses reflected human responses. Details are provided in Appendix B.

## Experiments

### Study 1

**Data:**
The dataset used in this study was collected by the Asociación para la Investigación de Medios de Comunicación (AIMC) in early 2017, from 1,511 Spanish internet users. Respondents provided demographic information on age and gender; they responded to survey items using a 1 to 7 Likert scale. The survey also collected data on their usage of ad-blockers. The survey aimed to investigate the relationships among seven latent variables related to online advertising:

| Path | Human | Baseline prompt | Demo prompt | LLM-Mirror | Omni prompt |
|---|---|---|---|---|---|
| Pleasure | 0.3359*** (0.0272) | 0.0328 (0.0620) | -0.0824*** (0.0418) | 0.3560*** (0.0159) | 0.5100*** (0.0272) |
| Credibility | 0.2559*** (0.0223) | -0.0112 (0.0725) | -0.1101*** (0.0285) | 0.1700*** (0.0138) | 0.2510*** (0.0146) |
| Economic | 0.2507*** (0.0220) | -0.0153 (0.0636) | 0.0204 (0.0638) | 0.4750*** (0.0145) | 0.3030*** (0.0145) |
| Intrusiveness | -0.1641*** (0.0223) | -0.0016 (0.0591) | -0.0432 (0.1233) | -0.1480*** (0.0122) | -0.1370*** (0.0136) |
| Clutter | -0.0695*** (0.0212) | 0.0257 (0.0452) | 0.0104 (0.0314) | -0.1280*** (0.0116) | -0.0730*** (0.0120) |

Table 2: Path Coefficients for Study 1(Path to *Attitude*)
Standard deviation in parentheses. p < 0.1;* p < 0.5;** p<0.01;***

knowledge of ad blockers (knowledge), attitude toward it (attitude), pleasure induced by it (pleasure), perceived credibility (credibility), economic evaluation (economic), perceived intrusiveness (intrusiveness), and perceived clutter (clutter). These relationships and their influence on ad-blocker usage were analyzed using PLS-SEM (Redondo and Aznar 2018). The specific survey items and the latent variables they measure are detailed in Appendix C.

**Experimental Design:**
We focused on generating survey responses using LLM to replicate the relationships among six latent variables: pleasure, credibility, economic evaluation, intrusiveness, clutter, and attitude. Specifically, we examined how the first five variables influence attitude. Among these variables, economic evaluation was measured with three questions, while pleasure, credibility, intrusiveness, clutter, and attitude were each measured with four questions. In our approach, we provided the LLM with a respondent's answers for the five latent factors as prompts. Based on these inputs, the LLM generated responses to the questions measuring attitude. Study 1 aims to evaluate how well the LLM can replicate human responses and investigate whether LLM-generated responses align with the hypotheses of the conceptual model.

**Results:**
Table 2 presents the results of PLS-SEM conducted using human responses and LLM-generated responses. The LLM prompt conditions, including the Omni prompting approach, are described in Table 1. The results in Table2 show that the path coefficients in both the Omni prompting and the LLM-Mirror approaches closely resemble those derived from human responses. For all paths, significant estimates were consistent with human responses, and the direction and magnitude of the coefficients were also closely aligned. With the Baseline prompting and the Demo prompting approaches, the majority of path coefficients were diverged from human response results, underscoring that providing prior questions and responses significantly improves the LLM's ability to mirror human responses.

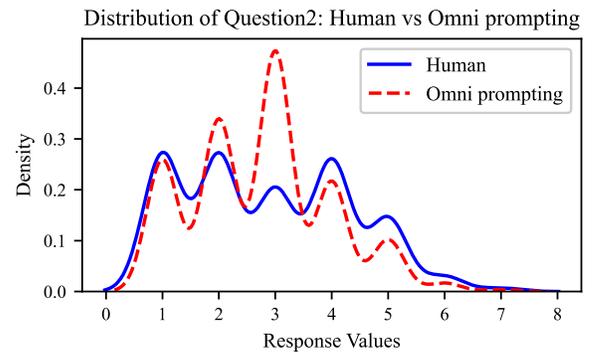

Figure 2: Study 1 KDE distribution of the Omni prompting

Figure 2 visualizes the response distribution for the Omni prompting approach. Consistent with the findings from the PLS-SEM analysis, the KDE distributions also demonstrate that providing response information about latent factors enables LLM to effectively reflect the respondent's perspective.

The Jensen-Shannon divergence in Table 3 and the Wasserstein distance in Table 4 demonstrate that the LLM responses in the Omni prompting and the LLM-Mirror approaches closely follow human responses. Consistent with previous results, providing prior responses leads to noticeable improvements compared to conditions without prior information.

Study 1 shows that providing prior questions and answers significantly enhances LLM's ability to reflect respondents. The improvement is also maintained in the LLM-Mirror approach. Additionally, path coefficients from the LLM responses closely align with those from human responses, suggesting the potential applicability of LLM-generated responses for hypothesis pre-testing.

|      | Baseline prompt | Demo prompt | LLM-Mirror | Omni prompt |
|------|-----------------|-------------|------------|-------------|
| Q1   | 0.5360          | 0.2770      | 0.0537     | 0.0780      |
| Q2   | 0.2534          | 0.2418      | 0.0710     | 0.0332      |
| Q3   | 0.4103          | 0.3453      | 0.1048     | 0.0806      |
| Q4   | 0.4303          | 0.3531      | 0.0870     | 0.0604      |
| Mean | 0.4075          | 0.3043      | 0.0791     | 0.0631      |

Table 3: Jensen-Shannon Divergence for each approach

|      | Baseline prompt | Demo prompt | LLM-Mirror | Omni prompt |
|------|-----------------|-------------|------------|-------------|
| Q1   | 1.1621          | 0.7240      | 0.2932     | 0.6453      |
| Q2   | 0.7459          | 0.7611      | 0.3891     | 0.2991      |
| Q3   | 1.0754          | 0.7333      | 0.7148     | 0.7750      |
| Q4   | 0.9755          | 0.7584      | 0.6896     | 0.5387      |
| Mean | 0.9897          | 0.7442      | 0.5522     | 0.5645      |

Table 4: Wasserstein distance for each approach

**Study 2**

**Data:**
To assess the generalizability of the findings from Study 1 to another dataset, additional experiments were conducted using the data from Damberg, Schwaiger, and Ringle (2022). The data, collected via a commercial German market research institute using quota sampling, consisted of mandatory Likert scale responses from 675 German cooperative bank customers. Respondents provided demographic information on age group, gender, marital status, education, occupational status, and monthly income. The survey in the study aimed to find the relationships among corporate reputation (COMP: perceived competence; LIKE: perceived likeability), relational trust (TRUST), customer satisfaction (SAT), and customer loyalty (LOY) using PLS-SEM. It specifically investigated the mediating roles of relational trust and customer satisfaction in the relationship between corporate reputation and customer loyalty. Corporate reputation was measured with two questions for likeability and three questions for competence, satisfaction with three questions, trust with four questions, and loyalty with three questions. The specific survey items and the latent variables they measure are detailed in Appendix D.

**Experimental Design:**
The study consisted of two parts. As shown in Figure 3, Case 1 explored the relationships between loyalty and four latent variables: competence, likeability, relational trust, and satisfaction. By providing the LLM with prior responses of four latent variables to answer three questions related to loyalty, Case 1 was structured as an extension of Study 1. Case 1 aimed to demonstrate the potential for LLMs to be generalized as effective tools for pre-testing hypotheses by applying them to a different context.

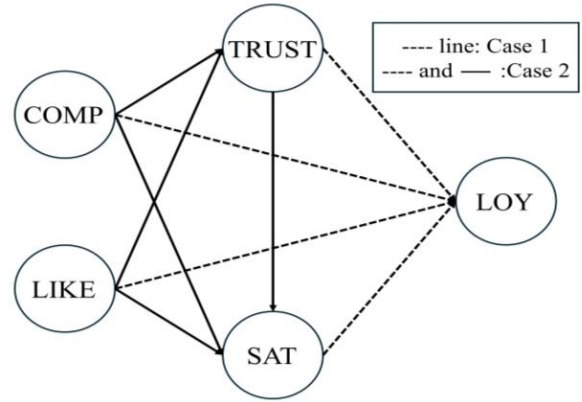

Figure 3: Fraction of Theoretical model Adapted from Damberg, Schwaiger, and Ringle (2022), Figure 1, page 3

In Case 2, as illustrated in Figure 3, the focus shifted to two latent variables, competence and likeability, to examine how trust and satisfaction mediate the relationship between corporate reputation and loyalty, as demonstrated in Damberg, Schwaiger, and Ringle (2022). Case 2 was designed to evaluate whether the LLM can capture more complex relationships by feeding LLM with the prior responses related to competence and likeability to capture the relationships among five latent variables.

**Case 1:**
The purpose of the first experiment was to confirm whether the findings from Study 1 could be applied to a different context. Two approaches—Omni prompting and Demo prompting— were compared, as outlined in Table 1. The results of the analysis, conducted using PLS-SEM, are presented in Table 5.

|                   | Human      | Demo prompt | Omni prompt |
|-------------------|------------|-------------|-------------|
| LIKE→ LOYALTY     | 0.1623*** (0.0510) | 0.0730 (0.0828) | 0.0647* (0.0340) |
| COMP→ LOYALTY     | 0.0157 (0.0519) | -0.0069 (0.0824) | 0.0154 (0.0376) |
| TRUST→ LOYALTY    | 0.3973*** (0.0649) | 0.0012 (0.0815) | 0.5324*** (0.0472) |
| SAT→ LOYALTY      | 0.2197*** (0.0667) | -0.1219 (0.0949) | 0.3221*** (0.0484) |

Table 5: Path Coefficients for Case 1

| Path | Human | Baseline prompt | Demo prompt | LLM-Mirror | Omni prompt |
|---|---|---|---|---|---|
| LIKE → TRUST | 0.3908*** (0.0389) | -0.0384 (0.0587) | 0.1095* (0.0595) | 0.3878*** (0.0286) | 0.3279*** (0.0286) |
| LIKE → SAT | 0.3596*** (0.0356) | -0.0123 (0.0511) | 0.0045 (0.0377) | 0.2458*** (0.0232) | 0.2676*** (0.0271) |
| LIKE → LOY | 0.1623*** (0.0510) | -0.0513 (0.0487) | -0.0361 (0.0404) | -0.1098*** (0.0294) | -0.0641*** (0.0256) |
| COMP → TRUST | 0.5037*** (0.0370) | -0.0097 (0.0649) | -0.0530 (0.0613) | 0.5215*** (0.0270) | 0.6438*** (0.0263) |
| COMP → SAT | 0.1616*** (0.0413) | -0.0146 (0.0577) | 0.0432 (0.0410) | 0.0020 (0.0249) | 0.2201*** (0.0476) |
| COMP → LOY | 0.0157 (0.0519) | 0.0295 (0.0547) | 0.0094 (0.0417) | 0.1724*** (0.0269) | 0.0750** (0.0323) |
| TRUST → SAT | 0.4432*** (0.0399) | 0.4156*** (0.0412) | 0.7306*** (0.0164) | 0.7375*** (0.0252) | 0.5076*** (0.0447) |
| TRUST → LOY | 0.3973*** (0.0649) | 0.0637 (0.0430) | 0.0026 (0.0339) | 0.0665 (0.0431) | 0.2089*** (0.0404) |
| SAT → LOY | 0.2197*** (0.0667) | 0.4884*** (0.0476) | 0.6960*** (0.0374) | 0.8060*** (0.0447) | 0.7138*** (0.0380) |

Table 6: Path Coefficients for Case 2

Based on the PLS-SEM results presented in Table 5, the findings showed that when prior information about the four latent variables of respondents was provided to the LLM for generating responses, it effectively replicated the relationship with loyalty. Notably, the coefficient of competence to loyalty was identified as insignificant in the real survey, and the LLM was able to capture this detail based on its responses.

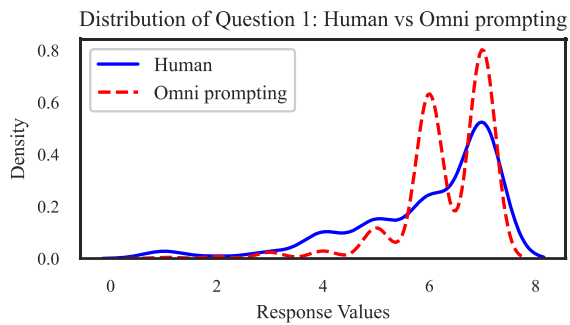

Figure 4: Study 2, Case 1 KDE distribution of the Omni prompting

Additionally, the experiment examined how closely the response distributions of the Omni prompting approach matched to those of humans by visually analyzing the KDE distribution (Figure 4) and quantitatively comparing those using the Jensen-Shannon divergence (Table 7) and the Wasserstein distance (Table 8).

Overall, the findings confirm that providing prior information to the LLM results in a closer alignment with original distribution.

| | Demo prompt | Omni prompt |
|---|---|---|
| Q1 | 0.2943 | 0.0551 |
| Q2 | 0.3526 | 0.0696 |
| Q3 | 0.1555 | 0.1894 |
| Mean | 0.2674 | 0.1047 |

Table 7: Jensen-Shannon Divergence for Demo prompting and Omni prompting

| | Demo prompt | Omni prompt |
|---|---|---|
| Q1 | 0.7748 | 0.3733 |
| Q2 | 1.3926 | 0.4563 |
| Q3 | 0.6830 | 1.2281 |
| Mean | 0.9501 | 0.6859 |

Table 8: Wasserstein Distance for Demo prompting and Omni prompting

**Case 2:**
Case 2 follows the same procedure but employs a more complex structural model. The prior information provided for each of the four approaches is detailed in Table 1. We compared the results with those from the original literature to evaluate how effectively the LLM replicated the findings of previous study. The path coefficients for PLS-SEM in Case 2 are presented in Table 6.

In the complex structural model, the Omni prompting approach successfully aligned with most of the path coefficients in terms of direction and magnitude. However, it exhibited limitations in capturing the insignificant coefficient of the path from competence to loyalty and incorrectly estimated the coefficient from likeability to loyalty. The LLM-Mirror approach faced similar problems, including the insignificant path from competence to satisfaction. One of the LLM-Mirror KDE distributions is provided below in Figure 5.

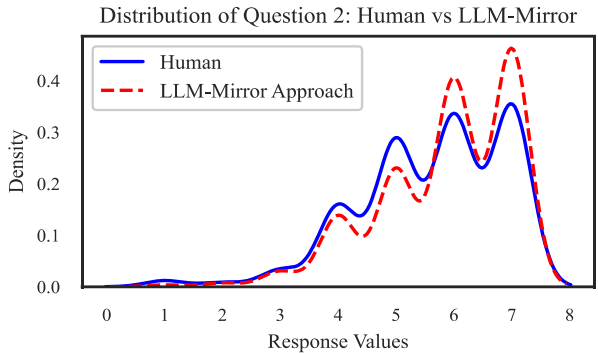

Figure 5: Study 2, Case 2 KDE distribution of the LLM-Mirror

As shown in the KDE distribution in Figure 5, the LLM-Mirror approach closely follows the real human distribution. The Jensen-Shannon divergence (Table 9) and the Wasserstein distance (Table 10), though exhibiting slightly lower similarity than those in the Omni prompting approach, still closely aligned with human distributions and showed notable differences compared to the Baseline prompting and the Demo prompting approaches.

|  | Baseline prompt | Demo prompt | LLM-Mirror | Omni prompt |
| --- | --- | --- | --- | --- |
| Mean | 0.3147 | 0.2417 | 0.0673 | 0.0556 |

Table 9: Average Jensen-Shannon Divergence for each approach

|  | Baseline prompt | Demo prompt | LLM-Mirror | Omni prompt |
| --- | --- | --- | --- | --- |
| Mean | 0.9239 | 0.9139 | 0.4812 | 0.4658 |

Table 10: Average Wasserstein Distance for each approach

The Jesen-Shannon divergence and Wasserstein distance calculated for each survey item are listed in Appendix A.

Our analysis identified the potential of the LLM-Mirror approach in Case 2 to capture the individual tendencies within a complex structural model.

Building on these findings, including the applicability of findings from Study 1 to other contexts, our study highlights the potential of LLMs to function as an LLM-Mirror by forming and utilizing user personas. These capabilities suggest their value as a pre-testing tool for hypothesis validation in survey-based research.

## Conclusion

In our study, we not only examined distributional similarities through metrics such as Jensen-Shannon divergence and Wasserstein distance but also compared the latent variable relationships in responses from humans and the LLM. By analyzing these relationships with path coefficients in structural models, we demonstrated that LLM can capture the underlying mechanisms of survey responses. Our findings expand the scope of LLM research from merely following response distributions to understanding relational dynamics, showing potential to capture both aggregate-level patterns and individual-level behaviors. These results underscore meaningful applications for LLMs in social science research.

Additionally, the LLM demonstrated its ability to follow original tendencies using the LLM-Mirror approach where user personas were formed based on respondents' given information. Our findings shed light on the possibility of using LLMs to generate persona and function as the LLM-Mirror for pre-testing hypotheses, even in scenarios with limited information about respondents.

## Limitation

Through Study 1 and Case 1 in Study 2, we found that LLM(GPT-4o) could generate human-like responses by prompting, supported by statistical metrics and path coefficient. However, Case 2 in Study 2, involving more complex latent variable relationships, showed some limitations in following certain tendencies, likely due to restricted access to GPT-4o. Fine-tuning, as shown by Gao et al. (2024), is expected to improve the model's alignment with human tendencies for complex relationships.

Our study explored the use of the LLM-Mirror for pre-testing or robustness checks, but was limited to online advertising and banking contexts with PLS-SEM. Future work is needed to expand these contexts and validate broader applications in social science, including testing with methods such as Regression, and ANOVA.

# Appendices

## Appendix A

|  | Baseline prompt | Demo prompt | LLM-Mirror | Omni prompt |
|---|---|---|---|---|
| Q1 | 0.7748 | 0.8311 | 0.2163 | 0.2667 |
| Q2 | 0.7985 | 0.5970 | 0.2326 | 0.2741 |
| Q3 | 1.1748 | 1.1259 | 0.3985 | 0.3719 |
| Q4 | 0.7881 | 0.7778 | 0.4607 | 0.4770 |
| Q5 | 0.8711 | 1.1081 | 0.5585 | 0.5585 |
| Q6 | 1.2059 | 0.5911 | 0.9511 | 0.9289 |
| Q7 | 1.1185 | 1.2696 | 0.8296 | 0.6919 |
| Q8 | 1.0607 | 1.1600 | 0.4993 | 0.4252 |
| Q9 | 0.7467 | 0.7481 | 0.2563 | 0.3407 |
| Q10 | 0.6993 | 0.9304 | 0.4089 | 0.3230 |

Table 11: Wasserstein distance for Case 2 of Study 2

|  | Baseline prompt | Demo prompt | LLM-Mirror | Omni prompt |
|---|---|---|---|---|
| Q1 | 0.2908 | 0.2602 | 0.0185 | 0.0225 |
| Q2 | 0.2989 | 0.2042 | 0.0091 | 0.0234 |
| Q3 | 0.4807 | 0.3197 | 0.0286 | 0.0477 |
| Q4 | 0.2774 | 0.2309 | 0.0428 | 0.0564 |
| Q5 | 0.3330 | 0.2389 | 0.1111 | 0.0868 |
| Q6 | 0.3775 | 0.1511 | 0.1672 | 0.1438 |
| Q7 | 0.2979 | 0.3238 | 0.1308 | 0.0715 |
| Q8 | 0.3228 | 0.3036 | 0.0774 | 0.0363 |
| Q9 | 0.2383 | 0.1658 | 0.0291 | 0.0437 |
| Q10 | 0.2307 | 0.2187 | 0.0581 | 0.0234 |

Table 12: Jensen-Shannon Divergence for Case 2 of Study 2

# Appendix B

|  | Baseline prompt | Demo prompt | LLM-Mirror | Omni prompt |
|---|---|---|---|---|
| Mean | 62.92% | 59.46% | 69.42% | 71.38% |

Table 13: Consistent Analysis of Study1

|  | Demo prompt | Omni prompt |
|---|---|---|
| Mean | 51.51% | 64.54% |

Table 14: Consistent Analysis of Case 1 of Study 2

|  | Baseline prompt | Demo prompt | LLM-Mirror | Omni prompt |
|---|---|---|---|---|
| Mean | 59.59% | 52.56% | 70.89% | 73.01% |

Table 15: Consistent Analysis of Case 2 of Study 2

**Appendix C**

| Latent variables | Items |
|---|---|
| Attitude toward online advertising | I think Internet advertisements are worth it. |
| | Generally, I consider Internet advertising to be a good thing. |
| | My general opinion about Internet advertising is highly favorable. |
| | I appreciate seeing advertising messages on the Internet. |
| Pleasure induced by online advertising | Internet advertising is very entertaining. |
| | Sometimes I take pleasure in thinking about what I saw or heard on online ads. |
| | Viewing online advertisements is a pleasant experience for me. |
| | Sometimes online advertising is even more enjoyable than other Internet content. |
| Perceived credibility of online advertising | Consumers may obtain reliable information through Internet advertising. |
| | Most Internet advertisements are trustworthy. |
| | Online advertisements reliably inform about the quality of products. |
| | Internet advertisements accurately reflect what products are like. |
| Economic evaluation of online advertising | Internet advertising contributes to society's economic development. |
| | Internet advertising helps raise our standard of living. |
| | Online advertisements promote competition, which benefits consumers. |
| Perceived intrusiveness of online advertising | Online advertising gets in the way of my Internet searches. |
| | Online advertising disrupts my activity on the Internet. |
| | Online advertising distracts me from my objectives while on the Internet. |
| | Internet advertisements intrude on the content I am accessing. |
| Perceived online advertising clutter | There are too many advertisements on the Internet. |
| | Internet advertisements are very repetitive. |
| | Web sites are full of advertising messages. |
| | We Internet users are inundated with so much online advertising. |

Table 16: Survey items and corresponding latent variables measured in Redondo and Aznar (2018). Only the survey items selected for Study 1 are listed here, not the complete set.

**Appendix D**

| Latent variables | Items |
|---|---|
| LIKE (perceived likeability) | I can identify better with my main bank than with other banks. |
| | If my bank no longer existed, I would regret it more than with other banks. |
| COMP (perceived competence) | My main bank is a leading provider in the market. |
| | As far as I know, my main bank enjoys a good reputation. |
| | I believe that my bank provides services of the highest standard. |
| SAT (customer satisfaction) | My main bank meets my expectations. |
| | I have a positive attitude towards my main bank. |
| | I prefer my main bank to other banks. |
| LOY (customer loyalty) | How likely is it that you will remain a customer of your bank? |
| | I will purchase new banking products in the future. |
| | In the future, I will make use of other banking products or financial services offered by my bank. |
| TRUST (relational trust) | My main bank always listens to me when I share my concerns and problems. |
| | My main bank always responds to my concerns and problems with constructive solutions. |
| | My main bank and I share the same values. |
| | I have the feeling that my bank always acts in accordance with the wishes of its customers. |

Table 17: Survey items and corresponding latent variables measured in Damberg, Svenja, and Ringle (2023). Only the survey items selected for Study 2 are listed here, not the complete set.